\documentclass[pra,twoside,showpacs,superscriptaddress,twocolumn]{revtex4}

\usepackage[dvips]{graphics}

\begin{document}

\title{Quantum multimeters: A programmable state discriminator}

\author{Miloslav Du\v{s}ek}
\affiliation{Department of Optics, Palack\'y University,
     17.~listopadu 50, 772\,00 Olomouc, Czech~Republic}
\affiliation{Research Center for Quantum Information, Slovak Academy
     of Sciences, D\'{u}bravsk\'{a} cesta 9, 842\,28 Bratislava,
     Slovakia}

\author{Vladim{\'\i}r Bu\v{z}ek}
\affiliation{Research Center for Quantum Information, Slovak Academy
     of Sciences, D\'{u}bravsk\'{a} cesta 9, 842\,28 Bratislava,
     Slovakia}
\affiliation{Faculty of Infomatics, Masaryk University,
     Botanick\'{a} 68a, 602\,00 Brno, Czech~Republic}

\date{January 20, 2002}

\begin{abstract}
We discuss a possibility to build a programmable quantum measurement
device (a ``quantum multimeter''). That is, a device that would be able
to perform various desired generalized, positive operator value measure
(POVM) measurements depending on a quantum state of a ``program
register''. As an example, we present a ``universal state
discriminator''. It serves for the unambiguous discrimination of a pair
of known {\em non-orthogonal} states (from a certain set). If the two
states are changed the apparatus can be switched via the choice of the
program register to discriminate the new pair of states unambiguously.
The proper POVM is determined  by the state of an auxiliary quantum
system. The probability of successful discrimination is not optimal for
all  pairs of non-orthogonal states from the given set. However, for
some subsets it can be very close to the optimal value.
\end{abstract}


\maketitle


\section{Introduction}

Quantum measurements are inevitable parts of all quantum devices. They
represent the final step of any quantum computation \cite{N+Ch}. The
whole sets of quantum measurements are of vital importance  in the
quantum-state estimation \cite{helst,hol}. Therefore, there is a natural
question whether it is possible to construct a universal (multi-purpose)
quantum measurement device. That is, an apparatus that could perform
some large class of generalized measurements (POVM) in such a way that
each member of this class could be selected by a particular quantum
state of a ``program register''.

The generalized measurement is defined by the fact that the probability
of each of its result (the number of results may be, in general, larger
than the dimension of the Hilbert space of the system under
consideration) is given by the expression $p_\mu =
{\mathrm{Tr}}_{\mathrm{S}} \left( {\mathsf{A}}_{\mu} \rho_{\mathrm{S}}
\right)$, where $\rho_{\mathrm{S}}$ is the state of the system and
${\mathsf{A}}_{\mu}$ are \emph{positive} operators that constitute the
decomposition of the identity operator ($\sum_\mu {\mathsf{A}}_{\mu}
=\openone$). This is the reason why it is called Positive Operator
Valued Measure \cite{helst,hol,per}. Each POVM can be implemented using
an ancillary quantum system in a specific state and realizing a
projective von Neumann measurement on the composite system
\cite{neumark}. In other words,  if one has an ``input'' (measured)
state $\rho_{\mathrm{S}}$ in the Hilbert space
${\mathcal{H}}_{\mathrm{S}}$ it is always possible to find some state
$\rho_{\mathrm{A}}$ in a space ${\mathcal{H}}_{\mathrm{A}}$ and a set of
orthogonal projectors $\{{\mathsf{E}}_{\mu}\}$ acting on
${\mathcal{H}}_{\mathrm{S}} \otimes {\mathcal{H}}_{\mathrm{A}}$
($\sum_\mu {\mathsf{E}}_{\mu} =\openone$) such that
\begin{equation}
  {\mathsf{A}}_{\mu} = {\mathrm{Tr}}_{\mathrm{A}}
     \left( {\mathsf{E}}_{\mu} \rho_{\mathrm{A}} \right)
\label{defA}
\end{equation}
are  positive operators as discussed above.

In general, we can assume, that the initial state of the ancilla can be
prepared with an arbitrary precision. The ancilla can be considered as a
part of the ``program register''. Further, we note that the general
projection measurement on the composite system can be represented by a
unitary transformation on the composite system followed by a fixed
projection measurement (e.g., independent projective measurements on
individual qubits). Therefore the problem of designing the programmable
quantum multimeter reduces to the question whether an arbitrary unitary
operation (on the Hilbert space with a given dimension) can be encoded
in some quantum state of a program register of a finite dimension. It
was shown that the answer to this question is ``No''. Nielsen and Chuang
proved that any two inequivalent operations require orthogonal program
states \cite{U-psi}. Thus the number of encoded operations cannot be
higher than the dimension of the Hilbert space of the program register.
Since, in general, the set of all unitary operations can be infinite,
the result of Nielsen and Chuang implies that no universal programmable
gate array can be constructed using finite resources. They showed,
however, that if the gate array is probabilistic, a universal gate array
is possible. A probabilistic array is one that requires a measurement to
be made at the output of the program register, and the output of the
data register is only accepted if a particular result, or set of
results, is obtained. This will happen with a probability, which is less
than one. Vidal and Cirac \cite{V+C} have presented a probabilistic
programmable quantum gate array with a finite program register which can
realize a family of operations with one continuous parameter. Recently,
Hillery {\em et al.} \cite{qproc} have proposed more general quantum
processor that can perform probabilistically any operation (not only
unitary) on a qubit. Another aspect of  encoding  quantum operations in
states of a program register has been discussed by Huelga {\em et al.}
\cite{huelga}. They dealt with the so called teleportation of unitary
operations. Unfortunately, the probabilistic realization of unitary
operations cannot help to built a programmable quantum multimeter in the
way mentioned above. The reason is that the probabilistic implementation
of a given operation leads, at the end, to a \emph{different} POVM than
the deterministic implementation of the same operation would lead to.

In general, we can describe a ``quantum multimeter'' as a (fixed)
unitary operation acting on the measured system (or a ``data register'')
and an ancillary system (``program register'') together and a (fixed)
projective measurement realized afterwards on the same composite system.
Clearly, such a device can perform only a restricted set of POVM's. One
can, therefore, ask what is the optimal unitary transformation that
enables us to implement ``the largest set of POVM's'' (in comparison
with the set of POVM's that would be obtainable when we allowed any
unitary transformation on the same Hilbert space). One can also ask what
unitary transformation can help to approximate all the POVM's (generated
by an arbitrary unitary transformation) with the highest precision
(fidelity) on average. Clearly, the last task requires definition of the
distance measure between two POVM's. This is an interesting problem {\em
per se}, however, it goes far beyond the scope of our considerations
here. Both optimization problems mentioned above are rather non-trivial.
Moreover, the introduced scheme is perhaps too general from a practical
point of view. Therefore, in the present paper we will concentrate our
attention on a more specific cases and especially on the problem of
state discrimination.

\section{Implementation}

\begin{figure}[htb]
\smallskip
\centerline{\resizebox{0.7\hsize}{!}{\includegraphics*{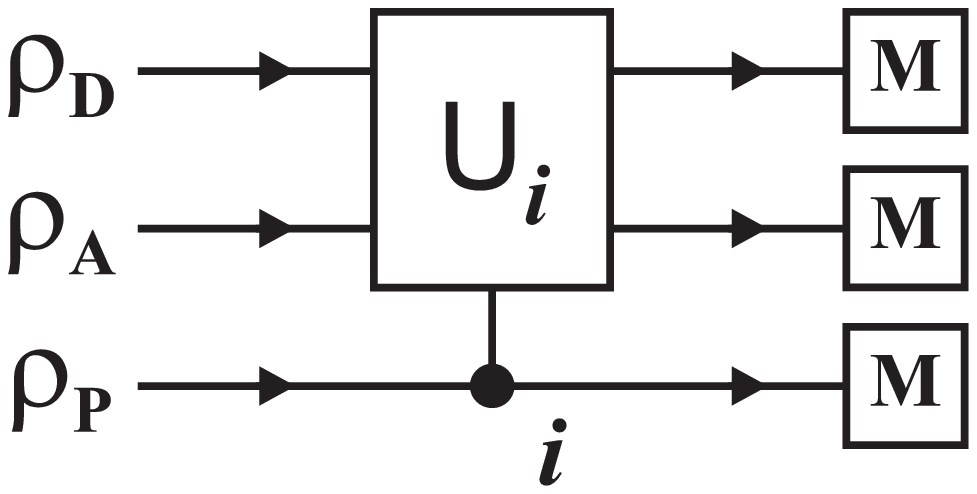}}}
\smallskip
\caption{A simple ``multi-purpose'' quantum measurement device (the
 quantum multimeter). The  total input state $\rho = \rho_{\mathrm{D}}
 \otimes \rho_{\mathrm{A}} \otimes  \rho_{\mathrm{P}}$ consists of the
 states of the data register, the ancilla, and the  program register.
 The ``program'' specifies the operation ${\mathsf{U}}_i$ applied on the
 data and the ancilla  [see Eq.~(\ref{processor})]. The final  step is a
 von Neumann  measurement realized separately on each  subsystem [see
 Eq.~(\ref{measurement1})].}
\label{fig1}
\end{figure}

Let us suppose that a ``customer'' asks our ``company'' to produce a
quantum measurement device that would be able to perform a finite number
of specific POVM's. Due the Neumark theorem \cite{neumark} we know that
each generalized measurement can be realized via a specific unitary
operation utilizing a certain state of the ancilla. Therefore we can
built an apparatus that is outlined in Fig.~\ref{fig1}. We choose a
large enough dimension of the ancillary system to cover the demands of
all particular POVM's and we built ``processors'' that provide a
particular required unitary operations. The device is programmed by the
state of the ancilla and by the proper choice of unitary transformation.
The selection of the unitary transformation is, in fact, a classical
procedure. Two inequivalent unitary operations correspond to two
orthogonal states of a program register \cite{U-psi}. However, even if
we restrict ourselves only to orthogonal program states addressing
exactly desired unitary operations we still obtain a large variety of
POVM's related to various possible states of ancilla. This is a ``side
effect'' that represents a ``bonus'' to our ``customer''. The set of
realizable POVM's can be further enlarged if one allows general program
states and a measurement on program register. (The question is only how
can such additional generalized measurements be useful for a
``customer''.)

Now let us analyze the scheme given in Fig.~\ref{fig1} in more detail.
The behavior of the device before a final measurement is performed can
be described by a unitary operation
\begin{equation}
   {\mathsf{P}} = \sum_{k=1}^{K} {\mathsf{U}}_k \otimes
   | {\mathrm{P}}_k \rangle \! \langle {\mathrm{P}}_k |,
\label{processor}
\end{equation}
where ${\mathsf{U}}_k$ are unitary operations acting on the
measured and ancillary systems together ($K$ is their total number)
and $| {\mathrm{P}}_k \rangle $ denote the orthonormal states of
a program register (the dimension of corresponding Hilbert space is
supposed to be $K$).  At this point we focus our attention to the final
measurement. We will assume independent projective measurements on
three subsystems (data, ancilla, and program) described by the
following set of projectors:
\begin{equation}
   {\mathsf{E}}_{ijk} = | {\mathrm{D}}_i \rangle \!
   \langle {\mathrm{D}}_i | \otimes | {\mathrm{A}}_j \rangle \!
   \langle {\mathrm{A}}_j | \otimes | {\mathrm{P}}'_k \rangle \!
   \langle {\mathrm{P}}'_k|
\label{measurement1}
\end{equation}
where vectors $| {\mathrm{D}}_i \rangle$ constitute an orthonormal basis
in the Hilbert space corresponding to the data register and $|
{\mathrm{A}}_j \rangle$ and $| {\mathrm{P}}'_k \rangle$ are orthonormal
bases in the spaces of ancilla and program, respectively. In general, $
\{ | {\mathrm{P}}_k \rangle \}_k$ and $\{ | {\mathrm{P}}'_k \rangle
\}_k$ represent two different bases. Direct application of
Eq.~(\ref{defA}) (we are tracing over both the ancilla and the program)
gives us a POVM (a set of positive operators) that depends on the states
of ancilla and program register:
\begin{eqnarray}
  {\mathsf{A}}_{ijk} &=&  \sum_{mnl}
  \langle {\mathrm{A}}_m | \, {\mathsf{U}}_n^{\dag}
  \Bigl[  | {\mathrm{D}}_i \rangle \! \langle {\mathrm{D}}_i |
  \otimes | {\mathrm{A}}_j \rangle \! \langle {\mathrm{A}}_j | \Bigr]
  {\mathsf{U}}_l \, \rho_{\mathrm{A}} | {\mathrm{A}}_m \rangle
        \nonumber \\  & \times &
  \langle {\mathrm{P}}_l | \rho_{\mathrm{P}} | {\mathrm{P}}_n \rangle
  \, \langle {\mathrm{P}}'_k | {\mathrm{P}}_l \rangle
  \langle {\mathrm{P}}_n | {\mathrm{P}}'_k \rangle.
\label{povm1}
\end{eqnarray}
Here $\rho_{\mathrm{A}}$ is the state of the ancilla and
$\rho_{\mathrm{P}}$ is the state of the program register.

If $| {\mathrm{P}}_k \rangle = | {\mathrm{P}}'_k \rangle $ for all
$k$, i.e., if the basis in Eq.~(\ref{processor}) is the same as the
basis of the measurement on the program register then
Eq.~(\ref{povm1}) simplifies:
\begin{eqnarray}
  {\mathsf{A}}_{ijk} &=&  \sum_{m} \langle {\mathrm{A}}_m |
  \, {\mathsf{U}}_k^{\dag}
  \Bigl [ | {\mathrm{D}}_i \rangle \! \langle {\mathrm{D}}_i |
  \otimes | {\mathrm{A}}_j \rangle \! \langle {\mathrm{A}}_j | \Bigr]
  {\mathsf{U}}_k \, \rho_{\mathrm{A}} | {\mathrm{A}}_m \rangle
    \nonumber \\
  & \times &  \langle {\mathrm{P}}_k | \rho_{\mathrm{P}} |
  {\mathrm{P}}_k \rangle.
\label{povm2}
\end{eqnarray}

For the sake of simplicity, let us first analyze the case when there
is no ancilla and unitary operations ${\mathsf{U}}_i$ concerns only
the data register (the measured system). Then Eq.~(\ref{povm2}) reads: 
\begin{equation}
  {\mathsf{A}}_{ik} =  {\mathsf{U}}_k^{\dag} | {\mathrm{D}}_i \rangle
  \! \langle {\mathrm{D}}_i |  {\mathsf{U}}_k \,
  \langle {\mathrm{P}}_k | \rho_{\mathrm{P}} | {\mathrm{P}}_k \rangle.
\label{povm3}
\end{equation}
Let us note that it depends only on the diagonal elements of the density
matrix describing the state of program register. The described situation
is, in fact, equivalent to the one when we are ``tossing a coin'' and,
depending on the {\em random} result, we are choosing one of several
projective tests.  In the given sense such an apparatus is equivalent to
the one that is ``programmed'' classically. Anyway, its input is quantum
and the program register can be set by the output of some other quantum
processor. For instance, in the case when the data input is represented
by a single spin-1/2 particle an apparatus of this kind can enable us to
make a ``software'' selection of the von Neumann measurement in one of
the three fixed orthogonal spatial axes (for this we would require a
three dimensional Hilbert space of the
program-register).\footnote{If the final measurement is the projection
   to the $z$-axis the required operations are the identity and two
   ``rotations'', $(\sigma_x + \sigma_z)/\sqrt{2}$ and   $(\sigma_y +
   \sigma_z)/\sqrt{2}$, where $\sigma$'s are   the Pauli matrices.}
Alternatively, we can perform any POVM that consists of a
``probabilistic mixture'' of these measurements in the sense explained
above. But the set of POVM's that can be realized using this scenario is
rather restricted. For instance, it is impossible to realize any
optimal unambiguous discrimination of two (known) non-orthogonal states
of spin-1/2 particle no matter which and how many unitary operations are
employed. To perform this task we need an ancilla.

However, before we leave this simple model let us look at the situation
when the basis $ \{ | {\mathrm{P}}_k \rangle \}_k$ is different from
the measurement basis $\{ | {\mathrm{P}}'_k \rangle \}_k$. If we
further assume the program register to be in a \emph{pure} state, such 
that $\langle {\mathrm{P}}_m | \rho_{\mathrm{P}} | {\mathrm{P}}_n
\rangle = \xi_m \xi_n^{*}$, then
\begin{equation}
  {\mathsf{A}}_{ik} =  {\mathsf{X}}_k^{\dag} \, | {\mathrm{D}}_i
  \rangle \!   \langle {\mathrm{D}}_i | \, {\mathsf{X}}_k,
\label{povm4}
\end{equation}
where the operators
$$
  {\mathsf{X}}_k = \sum_m {\mathsf{U}}_m \xi_m \langle {\mathrm{P}}'_k
  | {\mathrm{P}}_m \rangle
$$
are neither unitary nor Hermitean in general (this feature is related 
to the ``quantum processor'' proposed in Ref.~\cite{qproc}).

\section{Discrimination of quantum states}

In the following we will study a particular example of a ``quantum
multimeter'' serving for a programmable unambiguous state
discrimination. So, it is in place to say a few words about quantum
state discrimination now.

A general {\em unknown} quantum state cannot be determined completely by
a measurement performed on a single copy  of the system. But the
situation is different if {\em a priory} knowledge is available
\cite{helst,hol,per} -- e.g., if one works only with states from a
certain discrete set. Even quantum states that are mutually
non-orthogonal can be distinguished with a certain probability provided
they are linearly independent (for a review see Ref.~\cite{Chef}). There
are, in fact, two different optimal strategies \cite{barn-pha}: First,
the strategy that determines the state with the minimum probability for
the error \cite{helst,hol} and, second, unambiguous or error-free
discrimination (the measurement result never wrongly identify a state)
that allows the possibility of an inconclusive result (with a minimal
probability in the optimal case) \cite{usd-I,usd-D,usd-P,j+s,c+b}. We
will concentrate our attention to the unambiguous state discrimination.
It has been first investigated by Ivanovic \cite{usd-I} for the case of
two equally probable non-orthogonal states. Peres \cite{usd-P} solved
the problem of discrimination of two states in a formulation with POVM
measurement. Later Jaeger and Shimony \cite{j+s} extended the solution
to arbitrary a priori probabilities. Chefles and Barnett \cite{c+b} have
generalized Peres's solution to an arbitrary number of equally probable
states which are related by a symmetry transformation. Unambiguous state
discrimination were already realized experimentally. The first
experiment, designed for the discrimination of two linearly polarized
states of light, were done by Huttner \emph{et al.} \cite{discrim}. There are
also some newer proposals of optical implementations \cite{berg-pha}.
The interest in the quantum state discrimination is not only
``academic'' -- unambiguous state discrimination can be used, e.g., as
an efficient attack in quantum cryptography \cite{DJL}.

\section{``Universal'' discriminator}

Let us suppose that our ``customer'' wants to discriminate unambiguously
between two known non-orthogonal states. However, he/she would like to
have a possibility to ``switch'' the apparatus in order to be able to
work with several different pairs of states.

Let us have two (non-orthogonal) input states of a qubit. We can always
choose such a basis that they read $\alpha_0 | 0_{\mathrm{D}} \rangle
\pm \beta_0 | 1_{\mathrm{D}} \rangle$ with $\alpha_0 = \cos
(\varphi_0/2)$ and $\beta_0 = \sin (\varphi_0/2)$; the value of
$\varphi_0$ can be from $0$ to $\pi/2$ ($\varphi_0$ is the angle between
the two states). Let us have one additional ancillary qubit, initially
in a state $| 0_{\mathrm{A}} \rangle$. On both the ``data'' and the
ancilla we apply the following unitary transformation:
\begin{eqnarray}
 | 0_{\mathrm{D}} 0_{\mathrm{A}} \rangle  &\to&
    \cos \theta \, | 0_{\mathrm{D}} 0_{\mathrm{A}} \rangle +
    \sin \theta \, | 0_{\mathrm{D}} 1_{\mathrm{A}} \rangle,
    \nonumber \\
 | 1_{\mathrm{D}} 0_{\mathrm{A}} \rangle  &\to&
    | 1_{\mathrm{D}} 0_{\mathrm{A}} \rangle,
    \nonumber \\
 | 0_{\mathrm{D}} 1_{\mathrm{A}} \rangle  &\to&
    - \sin \theta \, | 0_{\mathrm{D}} 0_{\mathrm{A}} \rangle +
    \cos \theta \, | 0_{\mathrm{D}} 1_{\mathrm{A}} \rangle,
    \nonumber \\
 | 1_{\mathrm{D}} 1_{\mathrm{A}} \rangle  &\to&
    | 1_{\mathrm{D}} 1_{\mathrm{A}} \rangle,
\label{discU}
\end{eqnarray}
where $\cos \theta = \tan (\varphi_0/2)$. If we then make a von
Neumann measurement consisting of the projectors
${\mathsf{P}}_{+} = | + \rangle\!\langle + |$,
${\mathsf{P}}_{-} = | - \rangle\!\langle - |$, and
${\mathsf{P}}_{0} = \openone - {\mathsf{P}}_{+} - {\mathsf{P}}_{-}$,
where
\begin{equation}
 | \pm \rangle = \left( | 0_{\mathrm{D}} 0_{\mathrm{A}} \rangle
 \pm | 1_{\mathrm{D}} 0_{\mathrm{A}} \rangle\right) / \sqrt{2},
\label{measSt}
\end{equation}
we can unambiguously determine the input state (with a certain
probability of the success) \cite{discrim}. This measurement is optimal
in the sense that the probability of inconclusive result is the lowest
possible (and it is the same for both the states). The probability of
the successful discrimination is $2 \sin^2(\varphi_0/2)$ \cite{usd-P}.

\begin{figure}[thb]
\smallskip
 \centerline{\resizebox{0.45\hsize}{!}{\includegraphics*{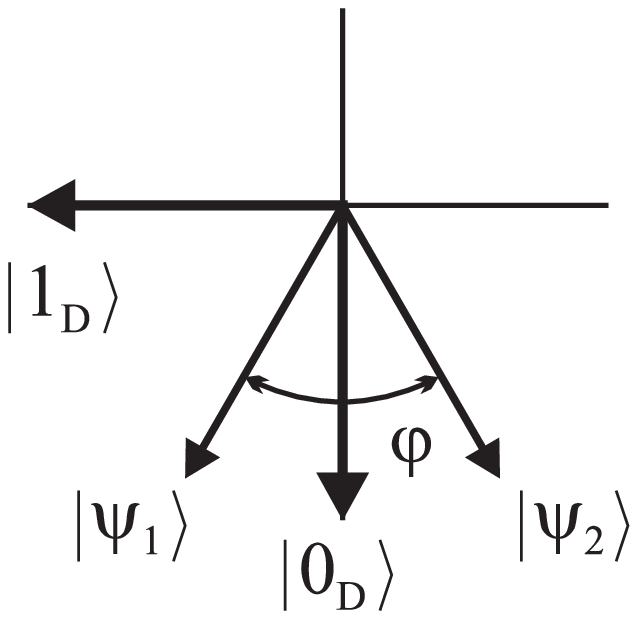}}}
\smallskip
\caption{The states $| \psi_1 \rangle$ and $| \psi_2 \rangle$ [defined
 by Eq.~(\ref{states})] with real coefficients $\alpha$ and $\beta$
 can be visualized in a two-dimensional real space. The angle
 $\varphi$ is related to the overlap of the two states: $\langle
 \phi_1 | \phi_2 \rangle = \cos \varphi = |\alpha|^2 + |\beta|^2 =
 2|\alpha|^2  - 1$.}
\label{fig2}
\end{figure}

Let us suppose now the set of pairs
\begin{eqnarray}
  | \psi_1 \rangle &=& \alpha \,| 0_{\mathrm{D}} \rangle
                      + \beta \,| 1_{\mathrm{D}} \rangle ,
  \nonumber \\
  | \psi_2 \rangle &=& \alpha \,| 0_{\mathrm{D}} \rangle
                      - \beta \,| 1_{\mathrm{D}} \rangle ,
\label{states}
\end{eqnarray}
where
$\alpha = \cos (\varphi/2)$ and $\beta = \sin (\varphi/2)$, for all
$\varphi$ from the interval $(0, \pi)$. That is, we consider all pairs
of states that lie on a real plane and that are located symmetrically
around the state $| 0_{\mathrm{D}} \rangle$; see Fig.~\ref{fig2}.
Further, let us suppose  that the ancillary qubit is allowed to be in an
arbitrary pure state
\begin{equation}
  a | 0_{\mathrm{A}} \rangle + b | 1_{\mathrm{A}} \rangle.
\label{anc}
\end{equation}
Thus the total input state reads
\begin{eqnarray}
\left(\, \alpha \,| 0_{\mathrm{D}} \rangle \pm
           \beta  \,| 1_{\mathrm{D}} \rangle \,\right)
  \otimes  \left(\, a \,| 0_{\mathrm{A}} \rangle +
                    b \,| 1_{\mathrm{A}} \rangle \,\right)
  = \nonumber \\
\alpha a \,| 0_{\mathrm{D}} 0_{\mathrm{A}} \rangle +
  \alpha b \,| 0_{\mathrm{D}} 1_{\mathrm{A}} \rangle \pm
  \beta  a \,| 1_{\mathrm{D}} 0_{\mathrm{A}} \rangle \pm
  \beta  b \,| 1_{\mathrm{D}} 1_{\mathrm{A}} \rangle .
\label{discIn}
\end{eqnarray}
After the action of transformation (\ref{discU}) on this state one
obtains the resulting state in the following form [the transformation
is fixed for all $\varphi$; still $\cos \theta = \tan (\varphi_0/2)$]
\begin{eqnarray}
&& \left( \alpha a \cos \theta - \alpha b \sin \theta \right)
   \,| 0_{\mathrm{D}} 0_{\mathrm{A}} \rangle +
   \nonumber \\
&& \left( \alpha a \sin \theta + \alpha b \cos \theta \right)
   \,| 0_{\mathrm{D}} 1_{\mathrm{A}} \rangle \pm
   \nonumber \\
&&\beta  a \,| 1_{\mathrm{D}} 0_{\mathrm{A}} \rangle \pm
  \beta  b \,| 1_{\mathrm{D}} 1_{\mathrm{A}} \rangle .
\label{discOut}
\end{eqnarray}
If the coefficients $a$ and $b$ in the state of the ancilla are chosen in
such a way that
\begin{equation}
   \left( \alpha a \cos \theta - \alpha b \sin \theta \right)
   = \beta a := q/\sqrt{2}
\label{cond}
\end{equation}
then the expression (\ref{discOut}) simplifies to the form
\begin{equation}
    q \,| \pm \rangle +
    {\mathrm{const}}_1
    \,| 0_{\mathrm{D}} 1_{\mathrm{A}} \rangle \pm
    {\mathrm{const}}_2
    \,| 1_{\mathrm{D}} 1_{\mathrm{A}} \rangle ,
\label{discSt}
\end{equation}
where the states $| \pm \rangle$ are defined by Eq.~(\ref{measSt}).
Clearly, applying the projective measurement introduced above one is
able to discriminate unambiguously states (\ref{states}) for any
given $\varphi \in (0, \pi)$ provided he/she has prepared the proper
state of the ancilla. The first term in Eq.~(\ref{discSt}) corresponds
to the successful discrimination, while the last two terms correspond to
inconclusive results. The probability of success is 
\begin{equation}
   P_{\mathrm{succ}} = |q|^2 =
   P_{\mathrm{opt}} R(\varphi, \varphi_0) =
   2 \sin^2 {\varphi \over 2} \, R(\varphi, \varphi_0),
\label{prob}
\end{equation}
where
\begin{equation}
  R(\varphi, \varphi_0)
  = \frac{\cos \varphi_0 \, (\cos \varphi + 1)}{1+
  \cos \varphi_0 - \sin \varphi \sin \varphi_0}
\label{R}
\end{equation}
is the ratio between the actual value of the probability of successful
discrimination and its optimal value. This expression is obtained from
the condition (\ref{cond}) together with the normalization relation
$|a|^2 + |b|^2 = 1$.

From above it follows that it is possible to implement a ``universal
quantum multimeter'' that is able to discriminate probabilistically but
unambiguously (with no errors) between two non-orthogonal states for
the large class of non-orthogonal pairs. The selection of the desired
regime (i.e., the selection of the pair of states that should be
unambiguously discriminate) is done by the change of the quantum state
of the ancillary qubit. The probability of the successful discrimination
can be optimal only for one such pair of
states.

In the limit case when $\varphi_0 = 0$, i.e. $\theta = \pi/2$ (this is
the fixed parameter of the employed unitary transformation), the
probability of the successful discrimination for different $\varphi$'s
(i.e., for different settings of the ancilla and different pairs of input
states) is the same as in the ``quasi-classical'' case,
$P_{\mathrm{succ}} = {1 \over 2} \sin^2 \varphi$. By a quasi-classical
approach we mean the probabilistic measurement when one randomly
selects\footnote{With the same probabilities provided that the
frequencies of the occurrence of the input states are also
    the same.}
the projective measurement in one of two orthogonal basis that both span
the two-dimensional space containing  both non-orthogonal states of
interest (\ref{states}). One basis consists of the state $| \psi_1
\rangle$ and its orthogonal complement $| \psi_1^\perp \rangle$. If one
finds the result corresponding to $| \psi_1^\perp \rangle$ he/she can be
sure that the state $| \psi_1 \rangle$ was not present. Analogously, the
other basis consists of the state $| \psi_2 \rangle$ and its orthogonal
complement.

On the other hand when $\varphi_0 = \pi/2$, i.e. $\theta = 0$, there
is no way how to fulfill the condition (\ref{cond}) with $a \ne 0$
(and $P_{\mathrm{succ}} \ne 0$) unless $\alpha = \beta = 1/\sqrt{2}$.
That is, only two orthogonal states (\ref{measSt}) can be unambiguously
discriminated.

\begin{figure}[htb]
\smallskip
 \centerline{\resizebox{\hsize}{!}{\includegraphics*{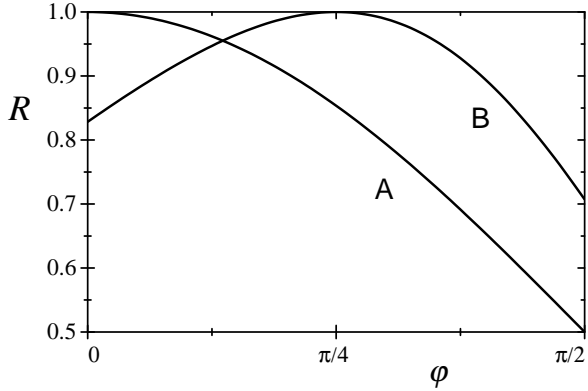}}}
\smallskip
\caption{The ratio $R(\varphi, \varphi_0)$ of the actual probability
 of successful discrimination to the optimal value of this probability
 as a function of the angle $\varphi$ between two considered state
 vectors. The curve (A) shows the ``quasi-classical'' limit
 ($\varphi_0=0$). The curve (B) represents the case when
 $\varphi_0=\pi/4$.}
\label{fig3}
\end{figure}

If the parameter $\varphi_0$ is somewhere in between 0 and $\pi/2$ the
probability of success (as a function of $\varphi$) is very close to
the optimal value in the relatively large vicinity of $\varphi_0$; see
Fig.~\ref{fig3}. However, for small values of $\varphi$ it goes below
the success probability of the quasi-classical case and for
$\varphi=\pi/2$ (orthogonal states) the probability of successful
discrimination is lower than unity.

One can ask for the optimal value of $\varphi_0$ in the sense that the
average probability of successful discrimination [or, alternatively,
function $R(\varphi, \varphi_0)$] over some chosen interval of
$\varphi$'s is maximal. For example, if we are interested in the
average value of $R(\varphi, \varphi_0)$ over the interval of
$\varphi$ from 0 to $\pi/2$ we find that it is maximized when
$\varphi_0 \approx 0.235\, \pi$ (the corresponding average value of
$R$ is $0.92$).

If our ``customer'' is not satisfied with such a device we can improve
its functionality using the scheme shown in Fig.~\ref{fig1}. By means
of a quantum state $|i\rangle$ of another ancillary system (program
register) one first selects the proper unitary transformation
${\mathsf{U}}_i$ that corresponds to the operation (\ref{discU}) with
$\varphi_0 = \varphi_i$ (the transformations differ only by
the values of $\varphi_0$ or $\theta$). This transformation should be
selected so that $\varphi_i$ is as close to the desired
$\varphi$ as possible. Then one prepares the state of the ancilla
(\ref{anc}) choosing the coefficients $a$ and $b$ such that the condition
(\ref{cond}) is fulfilled for the desired $\varphi$ and given $\varphi_0
= \varphi_i$.  This construction with ``switched operations'' enables us
to reach nearly the optimal probability of successful discrimination
for a large interval of angles $\varphi$.

\medskip

For pedagogical reasons till now we have only worked with the states
from a particular real subspace of the Hilbert space of the data qubit.
However, it should be stressed that the method works for any two
``input'' states that are symmetrically displaced with respect to $|
0_{\mathrm{D}} \rangle$. In other words, the condition (\ref{cond}) can
be fulfilled for any complex $\alpha$ and $\beta$. Simply,
$$
  \frac{b}{a} = \frac{1}{\sin \theta} \left( \cos \theta -
  \frac{\beta}{\alpha} \right).
$$
The probability of the successful discrimination of states then reads
\begin{equation}
   P_{\mathrm{succ}} = \frac{2 \sin \theta \, |\alpha \beta|^2}{1 -
   2 \cos \theta \, \Re (\alpha \beta)},
\label{genP}
\end{equation}
where $\Re (\alpha \beta)$ denotes a real part of $\alpha \beta$.

\section{Conclusions}

We have proposed a programmable quantum measurement device for the
error-free discrimination of two non-orthogonal states of qubit that
works with a large set of pairs of states. The device can be set to
discriminate unambiguously any two states that are symmetrically located
around some fixed state [in the sense of Eq.~(\ref{states})]. The
setting is done through the state of a program register that is
represented by another qubit. This means that the selection of the pair
of states to be discriminated unambiguously  depends on the state of a
program register. Two possible input states (of the ``data register'')
that are in correspondence with the program setting are never wrongly
identified but from time to time we can get an inconclusive result. The
probability of successful discrimination is optimal only for one program
setting. However, the device can be designed in such a way that the
probability of successful discrimination is very close to the optimal
value for a relatively large set of program settings.

We have also discussed some general questions concerning the
possibilities to built multi-purpose quantum measurement devices
(``quantum multimeters'') that could perform a required POVM depending
on a quantum state of their program register. Most of these questions
remain unanswered. For instance, let us suppose a set of all POVM's that
can be obtained if we combine the measured system with an ancilla of
some fixed dimension in an arbitrary state and carry out an arbitrary
projective (von Neumann) measurement on the composite system. This is
equivalent to carrying out an arbitrary unitary operation followed by
some fixed projective measurement. Imagine now that we  can change only
the state of the ancilla but our projective measurement (or unitary
transformation) is fixed. The question is: What measurement (operation)
do we need to approximate all the POVM's from the set introduced above
with the maximal average fidelity? Apparently, this question raises the
other interesting task: How to define the distance between two POVM's?
Such problems are not trivial, however, they open perspectives in
investigation of programmable quantum devices.


\begin{acknowledgments}

This research was supported under the European Union project EQUIP
(contract IST-1999-11053) and the project LN00A015 of the Ministry of
Education of the Czech Republic.

\end{acknowledgments}


\end{document}